\documentclass[usenatbib]{mn2e}

\usepackage{graphicx}
\usepackage{time}

\newcommand{\apj}{\mbox{\normalfont ApJ}}
\newcommand{\apjs}{\mbox{\normalfont ApJS}}
\newcommand{\apjl}{\mbox{\normalfont ApJL}}
\newcommand{\aj}{\mbox{\normalfont AJ}}
\newcommand{\mnras}{\mbox{\normalfont MNRAS}}
\newcommand{\aap}{\mbox{\normalfont A\&A}}

\newcommand{\nat}{\mbox{\normalfont Nature}}

\newcommand{\aapr}{\mbox{\normalfont A\&A~Rev.}}

\title[The MW without X]{The Milky Way without X: An alternative interpretation of the double red clump in the Galactic bulge}
\author[Y.-W. Lee, S.-J. Joo, and C. Chung]{Young-Wook Lee$^{}$\thanks{E-mail: ywlee2@yonsei.ac.kr}, Seok-Joo Joo$^{}$\thanks{E-mail: sjjoo@csa.yonsei.ac.kr}, and Chul Chung\\
$^{}$Center for Galaxy Evolution Research and Department of Astronomy, Yonsei University, Seoul 120-749, Korea}

\begin{document}

\date{Accepted 2015 August 22. Received 2015 August 7; in original form 2015 June 15}

\pagerange{\pageref{firstpage}--\pageref{lastpage}} \pubyear{2015}

\maketitle

\label{firstpage}

\begin{abstract}

The presence of two red clumps (RCs) in high latitude fields of the Milky Way bulge is interpreted as evidence for an X-shaped structure originated from the bar instability. 
Here we show, however, that this double RC phenomenon is more likely to be another manifestation of multiple populations observed in globular clusters (GCs) in the metal-rich regime. 
As in the bulge GC Terzan 5, the helium enhanced second generation stars in the classical bulge component of the Milky Way are placed on the bright RC, which is about 0.5 mag brighter than the normal RC originated from the first generation stars, producing the observed double RC. 
In a composite bulge, where a classical bulge can coexist with a boxy pseudo bulge, our models can also reproduce key observations, such as the dependence of the double RC feature on metallicity and Galactic latitude and longitude. 
If confirmed by Gaia trigonometric parallax distances, this would indicate that the Milky Way bar is not sufficiently buckled to form the X-shaped structure in the bulge, and suggest that the early-type galaxies would be similarly prevailed by super-helium-rich subpopulation.

\end{abstract}


\section{Introduction}

As the nearest early-type system, the Milky Way bulge provides a unique opportunity to study the details of its resolved stellar populations. 
Five years ago, the presence of double red clump (RC) was discovered in the higher latitude (${|b| > 5.5}$$^{\circ}$) fields of the Milky Way bulge from wide-field photometric surveys (\citealt{2010ApJ...724.1491M}; \citealt{2010ApJ...721L..28N}; {\citealt{2012A&A...544A.147S}}). 
These and follow-up observations have established that (1) the two RCs have almost identical mean colours, (2) the double RC feature is only evident among metal-rich stars, while metal-poor populations show only faint RC, (3) the separation between the two RCs is vanished at low Galactic latitudes, and (4) the relative strength of the two RCs changes strongly with longitude \citep{2012ApJ...756...22N, 2013MNRAS.432.2092N, 2015MNRAS.447.1535N, 2012A&A...546A..57U, 2014A&A...569A.103R}. 
These observations are widely accepted as evidence for the X-shaped bulge originated from the disk and bar instabilities, which led to the consensus that even higher latitude region of the Milky Way bulge has more characteristics of a ``pseudo bulge'', rather than a classical bulge \citep[][and references therein]{2010ApJ...724.1491M, 2011AJ....142...76S, 2012ApJ...757L...7L, 2013MNRAS.435.1874W, 2015MNRAS.447.1535N}.\footnote{Although the X-shaped structures are not uncommon among extragalactic bulges \citep[see, e.g.,][]{2006MNRAS.370..753B}, they are usually very faint and require significant image processing to reveal them, in contrast with the apparently strong signal claimed for the Milky Way.}
\vspace{0,1in}

Without the crucial distance information, however, an alternative interpretation, in which the brighter RC is representing an intrinsically brighter subpopulation rather than a distance effect, must be explored in detail to see whether it can equally reproduce the key observations described above. 
It is now well established that most globular clusters (GCs) host multiple populations with helium and light-elements enhanced second and third generation stars, for which the origin can be traced back to the chemical enrichments and pollutions by asymptotic-giant-branch (AGB) stars {and/or fast-rotating massive-stars (FRMS) \citep[][and references therein]{2004ApJ...611..871D, 2005ApJ...621L..57L, 2007A&A...464.1029D, 2009A&A...499..835V, 2009A&A...505..117C, 2012A&ARv..20...50G}. 
Some massive GCs, such as $\omega$~Cen and Terzan~5, also show evidence for supernovae enrichment \citep{1999Natur.402...55L, 2005ApJ...621..777P, 2009A&A...505.1099M, 2009Natur.462..480L, 2009Natur.462..483F, 2015ApJS..216...19L}. 
Chemical enrichments and pollutions similar to these might have affected stellar populations in the bulge as well. 
Furthermore, \citet{1994A&A...285L...5R} proposed that bulge populations might be helium-enriched due to their super-solar metallicity and the helium enrichment parameter ($\Delta Y/ \Delta Z$) between 2 and 3, with an average helium content of 0.31 -- 0.35.}

\begin{figure*}
\centering
\includegraphics[angle=0,scale=0.8]{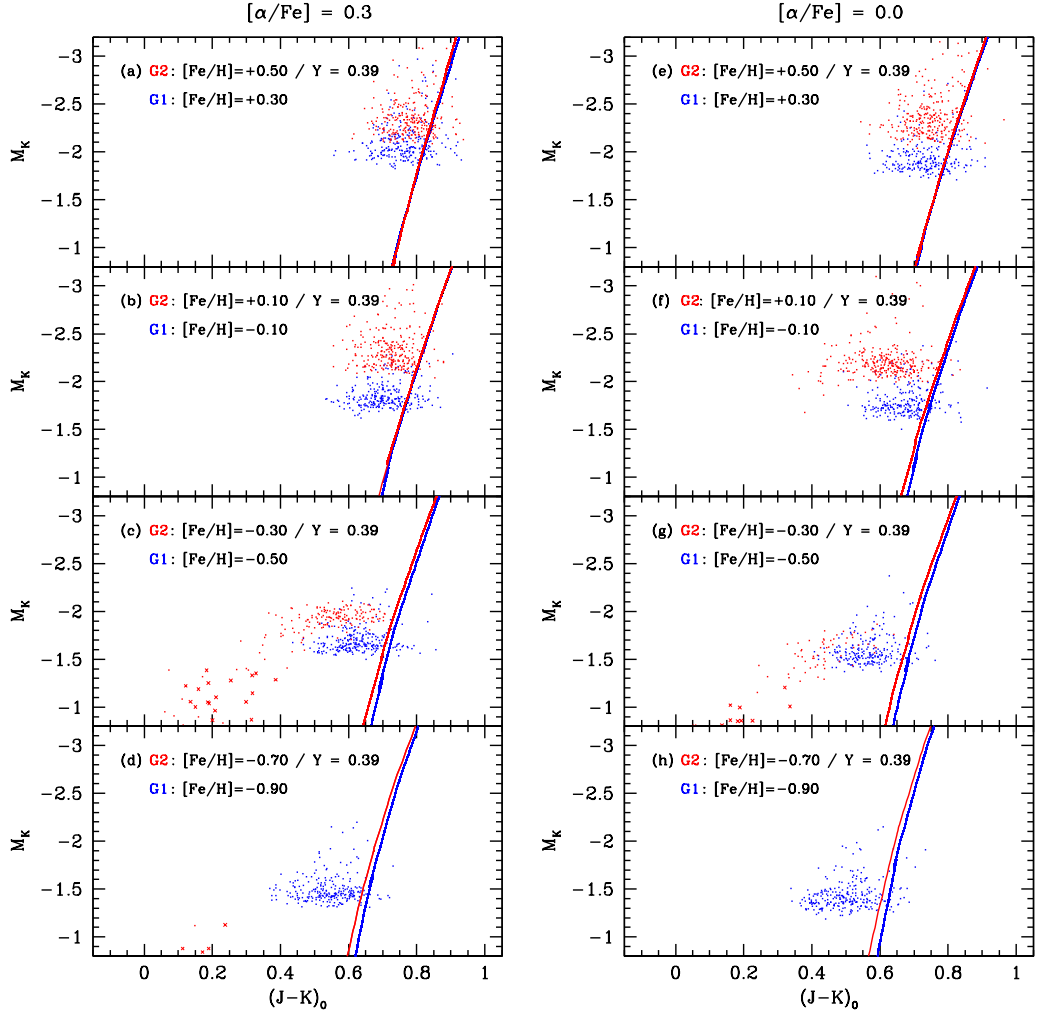}
\caption[]{Our population models for the two RCs from G1 and G2 in the ${(J-K, K)}$ CMDs. 
The models are presented at four different metallicity regimes to illustrate the sensitivity of the double RC feature on metallicity. 
The helium abundance for G1 follows the standard case (${\rm \Delta Y/\Delta Z} = 2$, ${\rm Y = 0.23 + 2Z}$), while that for G2 {is} adopted to be Y = 0.39.
{Panels (a) -- (d) are models for ${\rm [\alpha/Fe]} = 0.3$, while panels (e) -- (h) are for ${\rm [\alpha/Fe]} = 0.0$.} 
Crosses represent some RR Lyrae variables produced by G2.
}
\label{f1}
\end{figure*}

In general, super-helium-rich subpopulations are placed on the bluer horizontal-branch (HB) during the core helium-burning phase \citep{2005ApJ...621L..57L, 2012A&ARv..20...50G, 2013ApJ...762...36J, 2014MNRAS.437.1609M, 2014MNRAS.443L..15J}. 
However, when the metallicity is high enough as in the bulge, the strong metallicity effect overwhelms the helium effect, and the metal-rich counterpart would be placed on the brighter RC. 
This was correctly pointed out by \citet{2010ApJ...715L..63D} in their investigation of the double RC observed in the metal-rich bulge GC Terzan~5 \citep{2009Natur.462..483F}. 
The double RC observed in Terzan~5 is very analogous to that in the bulge in terms of the magnitude difference, metallicity, and population ratio. 
Both \citet{2010ApJ...724.1491M} and \citet{2010ApJ...721L..28N} noted this similarity, but they dismissed this possibility in the bulge mostly based on the fact that the two RCs in Terzan~5, unlike the double RC in the bulge, also show the difference in colour. 
The colour of the RC stars, however, can vary with metallicity, helium abundance, or age, similarly to the case of more metal-poor HB stars \citep[][see Figure~\ref{f1} below]{1994ApJ...423..248L}.

{Spectroscopic observations of red-giant-branch (RGB) and RC stars in the bulge field show a large range in metallicity, ${-1 \leq {\rm [Fe/H]} \leq 0.5}$, with a mean metallicity at almost solar \citep{2008A&A...486..177Z, 2010A&A...519A..77B, 2011A&A...534A..80H, 2013MNRAS.432.2092N, 2014A&A...569A.103R}. 
The metallicity distribution functions (MDFs) derived from these observations show bimodality with metal-rich and metal-poor populations. 
A metallicity gradient is also observed in the bulge along the minor axis, in the sense that the fraction of the metal-rich population is decreased at high Galactic latitudes. 
\citet{2011A&A...530A..54G} further showed that the $\alpha$-elements enhancement is decreased at higher metallicities \citep[see also][]{2012A&A...546A..57U, 2014AJ....148...67J}. 
Many of these observations and} recent theoretical studies have suggested a possibility of a composite bulge in the Milky Way, in which a classical bulge (CB) can coexist with a boxy pseudo bulge originated from the bar (\citealt{2010A&A...519A..77B}; \citealt{2011A&A...534A..80H}; \citealt{2015MNRAS.446.4039E}; {\citealt{2014A&A...569A.103R}; \citealt{2014A&A...562A..66Z}}; \citealt{2015arXiv150507048S}; \citealt{2015arXiv150504508S}).
In this composite bulge, low Galactic latitude ($|b| < 6^{\circ}$) fields would be dominated by the most metal-rich bar population, while relatively metal-poor CB population becomes more and more important at higher latitudes (\citealt{2010A&A...519A..77B}; {\citealt{2011A&A...530A..54G}}; \citealt{2013ApJ...763...26O}; {\citealt{2014A&A...569A.103R}}).
The purpose of this {paper} is to show that, in such a composite bulge, the multiple population models can equally explain the presence of the double RC and other key observations in the Milky Way bulge.

\section{Multiple population models for the double red clump}

The RC stars are metal-rich counterpart of core helium-burning HB stars in metal-poor GCs. 
In order {to study the RC populations} in the bulge, we have therefore constructed a series of synthetic HB models for the metal-rich populations, following the techniques developed by \citet{1994ApJ...423..248L} and as updated by \citet{2013ApJ...762...36J}. 
Our models are based on the Yonsei-Yale (${\rm Y^2}$) HB evolutionary tracks and isochrones with enhanced helium abundance \citep{2009gcgg.book...33H}, all constructed under the assumption that ${\rm [\alpha/Fe] = 0.3}$.
Following our recent investigations for the halo GCs with multiple populations \citep{2014MNRAS.443L..15J}, \citet{1977A&A....61..217R} mass-loss parameter $\eta$ was adopted to be 0.40, and the mass dispersion on the RC was adopted to be $\sigma_M = 0.010 M_\odot$ for each subpopulation.
\vspace{0,1in}

In our multiple population models for the CB component of the bulge, the faint RC (fRC) is produced by first-generation stars (G1), which were assumed to follow the standard helium enrichment parameter (${\rm \Delta Y/\Delta Z = 2}$, ${\rm Y = 0.23 + Z(\Delta Y/\Delta Z)}$), while the brighter RC (bRC) is populated by super-helium-rich (Y = 0.39) second-generation stars (G2). 
The choice of Y = 0.39 for G2 is based on the empirical constraint from super-helium-rich subpopulations in $\omega$~Cen \citep{2013ApJ...762...36J}, which is also comparable with the theoretical predictions for the helium content in the ejecta of both super-AGB stars and type II supernovae \citep{2013MNRAS.431.3642V, 1995ApJS..101..181W}. 
Note further that this choice of helium abundances for G1 and G2 (Y = 0.27 \& 0.39 at Z = 0.02), in the mean (Y = 0.33), is consistent with the suggestion from the analysis of the red giant-branch bump of the Milky Way bulge ({\citealt{1994A&A...285L...5R}}; \citealt{2013ApJ...766...77N}). 
Spectroscopy of RC stars in the bulge \citep{2011ApJ...732L..36D, 2012ApJ...756...22N, 2012A&A...546A..57U} indicates that stars in bRC are somewhat more metal-rich (0.12 -- 0.23 dex) than those in fRC, and therefore the metallicity difference between G2 and G1 was assumed to be $\Delta {\rm [Fe/H] = 0.2}$ dex in our models. 
The ages for G1 and G2 were adopted to be 12~Gyr and 10~Gyr, respectively \citep{2008ApJ...684.1110C}, and the two RCs are assumed to be equally populated \citep{2010ApJ...721L..28N}.
\vspace{0,1in}

Figure~\ref{f1} shows our models for the two RCs from G1 and G2 in the ${(J-K, K)}$ colour magnitude diagram (CMD). 
The models are presented at four different metallicity regimes relevant to the Milky Way bulge to illustrate the sensitivity of the double RC feature on metallicity. 
It is clear from the models in panel (b) that the separation in magnitude between the two RCs ($\sim$0.5 mag) is analogous to the observed difference, while the colour difference is negligible as is observed. 
More metal-poor models in panel (c) demonstrate, however, that, when the metallicity is decreased, the helium-rich G2 start to leave the RC regime to the bluer HB, while helium-normal G1 are placed on the fRC. 
When the metallicity is decreased further as in panel (d), all the core helium-burning stars produced by G2 are completely out of the bRC and are placed on the blue HB, while those from G1 are still on the fRC or red HB.
This case is qualitatively similar to the extended HB morphology observed in GCs with multiple populations such as NGC 2808 \citep[see][]{2005ApJ...621L..57L}. 
Consequently, no double RC would be observed in the relatively metal-poor population. 
This is mainly because helium-rich stars evolve more rapidly, and hence have lower masses at given age \citep[see][]{1994ApJ...423..248L, 2005ApJ...621L..57L}. 
Interestingly, the most metal-rich models in panel (a) show that the separation between the two RCs would be diminished or vanished in super-metal-rich (${\rm [Fe/H] \geq 0.2}$) population. 
This is because even G1, following the standard ${\rm \Delta Y/\Delta Z = 2}$, becomes very helium-rich (${\rm Y \approx 0.33}$) at super-metal-rich regime, increasing the luminosity of fRC.
In a two-component composite bulge, this most metal-rich regime would be more relevant to the bar component.

{The helium enhanced evolutionary tracks adopted in our modeling were constructed under the assumption that ${{\rm [\alpha/Fe]} = 0.3}$ for all ${{\rm [Fe/H]}}$ regimes. 
In the bulge fields, however, the ${\alpha}$-enhancement is observed to be decreased with increasing ${{\rm [Fe/H]}}$ \citep{2011A&A...530A..54G, 2012A&A...546A..57U, 2013MNRAS.432.2092N, 2014AJ....148...67J}. 
In order to investigate the effect of ${{\rm [\alpha/Fe]}}$, our ${\alpha}$-enhanced evolutionary tracks are rescaled back to ${{\rm [\alpha/Fe]}=0.0}$ by adopting the relation suggested by \citet{1993ApJ...414..580S}.
 Panels (e) -- (h) of Figure~\ref{f1} show our RC models constructed for ${{\rm [\alpha/Fe]}=0.0}$. 
At solar metallicity (${{\rm [Fe/H]} \approx 0.0}$), the scaled-solar (${{\rm [\alpha/Fe]}=0.0}$) models are fainter by ${\sim}$0.1 mag compared to the ${\alpha}$-enhanced (${{\rm [\alpha/Fe]}=0.3}$) models, but the magnitude difference between the two RCs is little affected. 
Furthermore, according to \citet[][see their Figure~19]{2013MNRAS.432.2092N}, the difference in ${{\rm [\alpha/Fe]}}$ between the two dominant components (at ${{\rm [Fe/H]} \approx  +0.15}$ and ${-0.25}$) in the bulge is observed to be only ${\sim}$0.1 dex, and therefore the effect of ${{\rm [\alpha/Fe]}}$ on the magnitude difference between the two RCs would be negligible.}

\begin{figure}
\includegraphics[angle=0,scale=0.65]{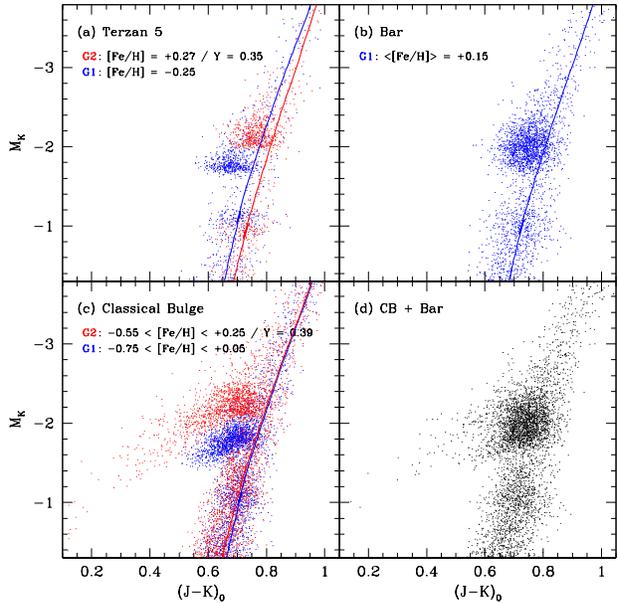}
\caption[]{
Same as Figure 1, but for the metal-rich GC Terzan~5 and the two components (bar and CB) in the Milky Way bulge. 
{Models for the bar and CB are superimposed in panel (d) to illustrate the composite bulge at low Galactic latitudes, where the population ratio between the bar and CB is adopted to be 2:1. Photometric errors, differential reddening and some depth effects ($\sigma_K \approx 0.047$, $\sigma_{J-K} \approx 0.033$ mag)} are included in our simulations.
}
\label{f2}
\end{figure}

\begin{figure}
\centering
\includegraphics[angle=0,scale=0.52]{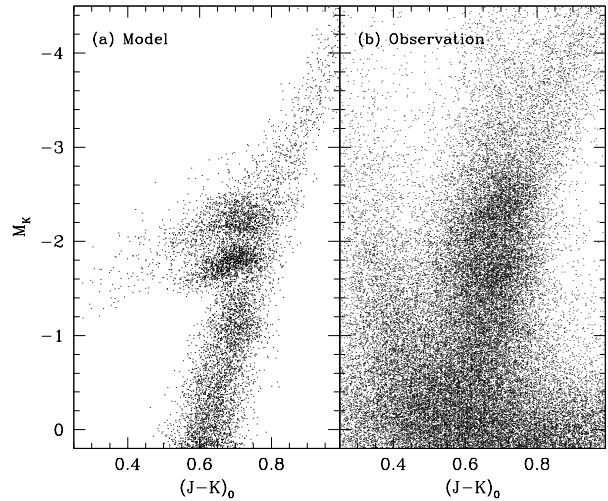}
\caption[]{
{Same as Figure 2, but our models for the CB are compared with the observed CMD for the high latitude ($b = -8^{\circ}$) field of the bulge (2MASS data from \citealt{2006AJ....131.1163S}; Figure 2 of \citealt{2010ApJ...724.1491M}). 
Note that the observed CMD is contaminated by the foreground disk stars at $(J-K)_0 \approx 0.4$ and 0.7.}
}
\label{f3}
\end{figure}

In {panel (c)} of Figure~\ref{f2}, our single metallicity models in Figure~\ref{f1} {(for ${{\rm [\alpha/Fe]}=0.3}$)} are combined to generate a composite model for the CB component of the bulge. 
For this, models in the metallicity range of $-0.75 \leq {\rm [Fe/H]} \leq +0.25$ were used, weighted according to the skewed Gaussian {MDF} truncated at both ends with a peak at ${\rm [Fe/H] = -0.1}$. 
In these models, the magnitude difference between the two RCs is $\Delta {K} = \Delta {I} \approx 0.43$ mag, while the colour difference is negligible, $\Delta {(J-K)} \approx 0.00$, which are in reasonable agreements with the observed differences {(see Figure~\ref{f3})}. 
Similar results are obtained even if the peak metallicity and age are varied by $\sim$$\pm 0.2$ dex and $\sim$$\pm 2$~Gyr.\footnote{Likewise, if a MDF with two Gaussian peaks at ${\rm [Fe/H]} = -0.25$ and $+0.15$ \citep{2010A&A...519A..77B, 2013MNRAS.432.2092N} is adopted roughly in the same metallicity range, the magnitude and colour differences are varied by only $\sim$0.02 and 0.00 mag, respectively.}
For comparison, panel (a) shows our model for the GC Terzan~5, where the metallicities for G1 and G2 were adopted from \citet{2011ApJ...726L..20O}, and the helium abundance for G2 was adopted to be Y = 0.35. 
Ages for G1 and G2 are identical to those adopted for the CB models (i.e., 12 \& 10~Gyr, respectively). 
Our models show that a larger difference in metal abundance ($\Delta {\rm [Fe/H]} = 0.5$ dex) between G2 and G1 in Terzan~5 is mainly responsible for the observed colour difference between the two RCs, while a smaller difference in helium abundance is also a contributing factor.
{If no difference in age is adopted between G1 and G2 for Terzan 5, as was assumed by \citet{2010ApJ...715L..63D}, the magnitude and colour differences are reduced by $\sim$0.05 and $\sim$0.01 mag, respectively.}
The panel (b) of Figure~\ref{f2} also presents our model for the bar component of the bulge, in which we assume that only metal-rich G1 (with an age of 10~Gyr) is populated considering the disk origin of the bar population. 
{Following the observed MDF for the most metal-rich bar component \citep{2010A&A...519A..77B, 2013MNRAS.432.2092N}, a Gaussian MDF with the mean ${\rm[Fe/H]} = +0.15$ and $\sigma_{\rm [Fe/H]} = 0.10$~dex is adopted.}
As described above, this {super-metal-rich G1}, following the standard ${{\rm \Delta Y/\Delta Z} = 2}$, becomes relatively helium-rich (${{\rm Y \approx }}$~0.31) and thus produces relatively bright (${{\rm M_K \approx} -2.0}$~mag) RC stars that would be placed between the bRC (G2) and fRC (G1) of the CB model in panel (c). 
{The panel (d) of Figure~\ref{f2} shows the CMD of the overlapping populations (i.e., CB + bar). 
In order to reproduce the low latitude fields of the bulge, the population ratio between the bar and CB is adopted to be 2:1. 
In this case, the split in the RC is not clear, but the Gaussian decomposition indicates two components separated by $\sim$0.30 mag, which is in reasonable agreement with the observed smaller difference at low Galactic latitudes, such as the Baade's window \citep[see Figure~7 of][]{2014A&A...565A...8C}.}

\begin{figure*}
\centering
\includegraphics[angle=0,scale=0.7]{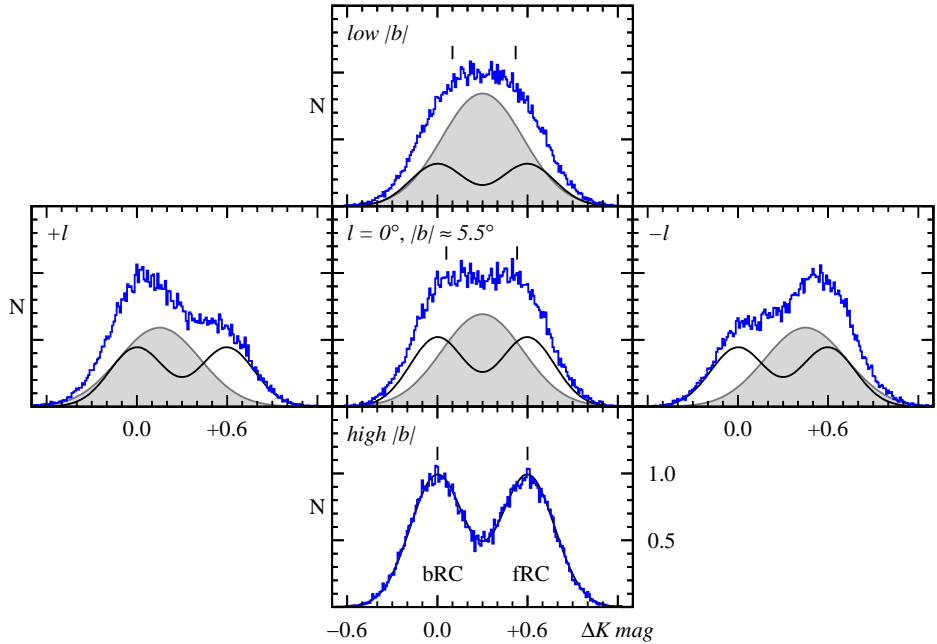}
\caption[]{
{Schematic diagrams illustrating the latitude (vertical panels) and longitude (horizontal panels) dependences of the double RC feature in a two-component bulge model. 
The mono-modal distribution in gray is for the bar component, while the bimodal distribution is for the CB component. 
The two components are assumed to be equally populated for three horizontal panels, while the fraction of the bar component decreases with increasing lbl in three vertical panels.}
Simulations for the sum of the two components (with $3 \times 10^4$ stars) are represented as blue histograms{, and the tick marks in vertical panels indicate the RC peak positions suggested from the Gaussian decomposition} (see the text).
}
\label{f4}
\end{figure*}

\vspace{0,1in}
Our models can reproduce all of the key observations described in Section 1. 
Firstly, as described above, the observed magnitude difference ($\sim$0.5 mag) between the two RCs is reproduced while maintaining no or little colour difference between them (see Figure~\ref{f2}). 
Secondly, the double RC feature is reproduced only for the relatively metal-rich (${\rm [Fe/H]} \geq -0.4$) populations as suggested from the observations \citep{2012ApJ...756...22N, 2012A&A...546A..57U}. 
Our models for the metal-poor populations in the bulge show only fRC (see Figure~\ref{f1} lower panel), because, unless very metal-rich (${\rm [Fe/H]} \geq -0.4$), super-helium-rich G2 cannot be trapped on the bRC. 
Thirdly, {as illustrated by a schematic diagram in Figure~\ref{f4} (three vertical panels),} the latitude dependence of the double RC feature can be explained within a two-component composite bulge. 
At low latitudes, the bulge is dominated by the most metal-rich bar population with only G1, which, on the RC, is placed between the two RCs from {a minority CB component, completely filling the gap of the double RC. 
At higher latitudes, however, contribution from the bar component is decreased/vanished, while} the CB component (with G1 and G2) becomes more and more important, {producing a larger magnitude difference between the two RCs with increasing $|b|$.} 
Finally, in a two-component bulge with a spheroidal CB embedded in a tilted bar (pseudo bulge), the longitude dependence of the RC luminosity function {can be reproduced in our model as well. 
As also illustrated in Figure~\ref{f4} (three horizontal panels)}, the observed luminosity function, in our scenario, is the sum of a CB component (with double RC) and a metal-rich bar component (with single RC). 
Towards negative longitudes the bar component is placed at the far side of the bulge and therefore fRC becomes gradually more populated, while the opposite effect is anticipated towards positive longitudes. 
{Most of these observations were obtained in $5^{\circ} < |b| < 6^{\circ}$ fields \citep[see Figure~11 of][]{2015MNRAS.447.1535N}, where the CB component is expected to be comparable to the bar component in terms of population ratio. 
More detailed population models that fully take into account the observed MDF and its dependence on Galactic latitude and longitude will be presented in our forthcoming paper (in preparation), where we will also compare the observed RC luminosity functions with the models constructed under various combinations of input parameters.}

\vspace{0,1in}
\section{Discussion}
The kinematics of RC and giant stars in the bulge can provide further constraints on our models.
The radial velocity ($V_r$) measurements show that the bulge rotates cylindrically \citep{2013MNRAS.432.2092N, 2014A&A...562A..66Z}, which is a strong evidence for the bar at low latitude fields ($|b| < 6$$^{\circ}$). 
For the higher latitudes, however, this is not necessarily inconsistent with our CB dominated scenario in this region of the bulge, because, for example, \citet{2015arXiv150504508S} showed that an initially non-rotating CB could absorb a significant fraction of the angular momentum from the bar within a few Gyr \citep[see also][]{2014A&A...562A..66Z}. 
The observed kinematics of stars in the two RCs are also consistent with our scenario. 
At $b = - 6^{\circ}$, \citet{2013A&A...555A..91V} found an excess of stars moving towards the Sun in the bRC, while that receding from the Sun is observed in the fRC.
In our composite bulge scenario, this can be understood by the depth ($\sim$2 kpc; $\Delta {\rm mag} \approx$ 0.5~mag) effect of the bar. 
Since the RCs from the bar and CB are mixed in the CMD of this field, RC stars in near side of the bar would be superimposed on bRC from CB, while those in far side of the bar is mixed with fRC from CB. 
Therefore, if the stars in the bar are in streaming motions \citep{2013A&A...555A..91V}, we would see some difference in $V_r$, in the mean, between bRC and fRC. 
At higher latitudes ($b = -8$ \& $-10^{\circ}$), however, there is no difference in $V_r$ between the two RCs \citep{2011ApJ...732L..36D, 2012A&A...546A..57U, 2014A&A...569A.103R}, which is consistent with our CB (with multiple populations) dominated scenario in these regions of the bulge.

\vspace{0,1in}
{One challenge to our multiple population model for the bulge is to understand the population ratio of G2 (He-enhanced) which is observed to be comparable to that of G1 (He-normal). 
In the metal-poor GCs, it is now well established that the formation of super-helium-rich subpopulation requires specific conditions in the central region of a massive proto-GC \citep[see. e.g.,][]{2008MNRAS.391..825D}, mainly to fulfill the apparently very extreme helium to metal enrichment (${\rm \Delta Y/\Delta Z} > 70$). 
When these GCs were disrupted to provide stars to halo field, the preferential removal of G1 would lead to the scarcity of G2 in the field. 
However, the situation is very different in the metal-rich (${\rm [Fe/H]} \approx 0.0$) system like the bulge, where the formation of super-helium-rich G2 requires only relatively modest enrichment parameter (${\rm \Delta Y/\Delta Z} \approx 5-6$). 
Helium enrichment similar to this magnitude is observed in the extragalactic HII regions \citep[see, e.g.,][]{1992MNRAS.255..325P}, and is also suggested from type II supernova models that form black holes above 20--25~$M_{\odot}$ \citep{1992A&A...264..105M, 1995ApJS...98..617T}. 
In addition to supernovae, AGB stars and FRMS would have also provided helium to G2, and therefore helium-rich G2 could have been formed ubiquitously in the metal-rich ``building blocks'' like Terzan~5. 
While the details of chemical evolution in the bulge and Terzan~5 require further investigations \citep[see][]{2010ApJ...715L..63D, 2012MNRAS.421L..44B}, it appears therefore that both G1 and G2 were effectively provided to the CB component of the Milky Way when merging and disruption of these building blocks were much more active in the early phase of the Milky Way formation.}

The distance measurements for the two RCs from soon to be released Gaia trigonometric parallax data would provide the most direct test for the suggested model. 
If the two RCs are due to the X-shaped structure, $\sim$0.5 mag difference in the distance modulus is expected between the bRC and fRC, while essentially no difference, in the mean, is predicted in our multiple population models. 
This should settle the issue once and for all. 
If confirmed, our result would suggest that the bar in the Milky Way bulge is not sufficiently buckled to form the X-shaped structure to be observed as double RC at higher latitudes ($|b| > 5.5^{\circ}$). 
Furthermore, if the split RC is mostly due to the effect of multiple populations, rather than a distance effect, the previous studies on the structure of the Milky Way bulge and bar based on RC, high latitude fields in particular, should be reexamined in the new paradigm. 
Finally, our result for the CB component of the Milky Way bulge, the nearest early-type system, would also suggest that the early-type galaxies would be similarly prevailed by super-helium-rich subpopulation. 
If $\sim$50\% of stars in early-type galaxies are helium and light elements enhanced G2, this would have profound impacts on the interpretation of integrated spectra by employing population synthesis models{, such as the origins of the Na enhanced giant elliptical galaxies \citep{2010Natur.468..940V, 2013ApJS..208....7J} and the UV upturn phenomenon \citep{2011ApJ...740L..45C}.}

\section*{Acknowledgments}

{We thank the referee for a number of helpful suggestions which led to several improvements in the manuscript. 
We also thank Sang-Il Han for his assistance in obtaining the observed data for Figure 3.}
Support for this work was provided by the National Research Foundation of Korea to the Center for Galaxy Evolution Research.

\bsp

\label{lastpage}

\end{document}